\documentclass[aps,twocolumn,reprint] {revtex4-1}
\usepackage{comment}
\usepackage{graphicx}
\usepackage{tabularx}
\usepackage{amssymb}
\usepackage{amsmath}

\usepackage[utf8]{inputenc}
\usepackage[T2A,T1]{fontenc}
\usepackage[english]{babel}

\usepackage{color}
\usepackage{ulem}
\usepackage{xcolor}
\usepackage{comment}

\usepackage[caption=false]{subfig}

\graphicspath{ {images/} }

\begin{document}

\title{Nanoparticle lattices with bases: Fourier modal method and dipole approximation}

\author{Ilia M. Fradkin}
\email{Ilia.Fradkin@skoltech.ru}
\affiliation{Skolkovo Institute of Science and Technology, Nobel Street 3, Moscow 143025, Russia}
\affiliation{Moscow Institute of Physics and Technology, Institutskiy pereulok 9, Moscow Region 141701, Russia}
\author{Sergey A. Dyakov}
\affiliation{Skolkovo Institute of Science and Technology, Nobel Street 3, Moscow 143025, Russia}
\author{Nikolay A. Gippius}
\affiliation{Skolkovo Institute of Science and Technology, Nobel Street 3, Moscow 143025, Russia}
\date{\today}

\begin{abstract}
The utilization of periodic structures such as photonic crystals and metasurfaces is common for light manipulation at nanoscales. One of the most widely used computational approaches to consider them and design effective optical devices is the Fourier modal method (FMM) based on Fourier decomposition of electromagnetic fields. Nevertheless, calculating of periodic structures with small inclusions is often a difficult task, since they induce lots of high-$k_\parallel$ harmonics that should be taken into account. In this paper, we consider small particle lattices with bases and construct their scattering matrices via discrete dipole approximation (DDA). Afterwards, these matrices are implemented in FMM for consideration of complicated layered structures.
We show the performance of the proposed hybrid approach by its application to a lattice, which routes left and right circularly polarized incident light to guided modes propagating in opposite directions. We also demonstrate its precision by spectra comparison with finite element method (FEM) calculations. The high speed and precision of this approach enable the calculation of angle-dependent spectra with very high resolution in a reasonable time, which allows resolving narrow lines unobservable by other methods.
\end{abstract}

\maketitle

\section{Introduction}

Electromagnetic metasurfaces representing subwavelength patterns on a chip are considered as one of the most effective and convenient tools for light manipulation on a chip. In particular, metasurfaces consisting of deep subwavelength nanoparticles are very convenient for the design of structures with desired properties. Indeed, small nanoparticles have a dipole response, which makes it easy to understand qualitatively the nature of the physical processes occurring in these structures. At the same time, the choice of dielectric or plasmonic material, the shape of the particle adjustment and the tuning of particle resonances give us wide opportunities in the determination of their optical properties for a demonstration of bright physical effects and their application for different purposes. 
Such metasurfaces have already been successfully implemented for holography \cite{zheng2015metasurface,ye2016spin,wei2017broadband}, demonstration and implementation of lattice plasmon resonances \cite{Augie2008,Rajeeva2018,Chu2008,kravets2008,kravets2018plasmonic,guo2017}, biosensors \cite{Shen2013,rodriguez2011,lodewijks2012}, spin-orbit coupling \cite{jiang2018metasurface} and many other purposes.

However, small particles, especially if they are plasmonic ones, induce high gradients of electromagnetic fields. This strongly complicates their numerical consideration by the means of Fourier modal method (FMM), which is the most common and natural approach for dealing with periodic structures. The fact is that FMM also known as Rigorous coupled-wave analysis (RCWA) is based on Fourier decomposition of electromagnetic fields and therefore requires to take into account too many harmonics to describe tiny inclusion, which makes it inefficient in this case. Several approaches partially solve this problem. Li's factorization rules help to solve the problem of concurrent jump discontinuities \cite{li1997new}. Adaptive spatial resolution \cite{granet2002parametric,weiss2009matched} technique also improves convergence rate for multiscale problems and particles of non-rectangular shape. However, for deep-subwavelength particles, an approach based on the consideration of the lattices in dipole approximation, which we proposed in our recent paper \cite{fradkin2019fourier,fradkin2018fourier} is more natural and fast than other ones. In that article, we considered only simple lattices, but obviously lattices with several particles in a cell pave the way for observation of a wide range of physical phenomena. Auxiliary variation of relative position and orientation of particles in a cell and the lattice itself makes it possible to obtain 2D crystals with diverse optical properties.

Here, we expand our approach based on the combination of discrete dipole approximation (DDA) with FMM \cite{fradkin2019fourier,fradkin2018fourier} for the consideration of lattices with bases (several particles in a cell in our case). We introduce the concept of the generalized effective polarizability tensor for several particles in a cell. An example of the method application is demonstrated in the paper: the lattice of two nanobars acts as a polarization-controlled grating coupler, which routes normally incident left- and right-hand circular polarization in opposite directions.

\section{Lattice with basis}

The general idea of the dipole consideration of lattice is to substitute the real particle with an ideal point electric dipole. Its dipole moment can be found by an application of polarizability tensor of this particle to the background electric field at the position of the particle:

\begin{equation}
    \mathbf{P}_{\beta,i}=\hat{\alpha}_\beta\mathbf{E}_{\beta,i}^\mathrm{bg}.
    \label{eq:1}
\end{equation}
Hereinafter, we use Latin letters $i$ and $j$ to enumerate the cells and Greek ones $\beta$ and $\gamma$ for particles in a cell. A convenient approach for a calculation of polarizability tensor, $\hat{\alpha}$, of a particle in an arbitrary environment, including one laying on the interface, is reported in \cite{fradkin2019fourier,fradkin2018fourier}.

In turn, the background electric field is the sum of the external electric field, $\mathbf{E}^0$, which would have been at the position of the particles for the structure without lattice,
and the field rescattered by all the neighboring particles in a lattice. We represent this quantity as a sum of contributions of the sublattice to which the considered particle belongs and all the other sublattices:

\begin{multline}
    \mathbf{E}^{\mathrm{bg}}_{\beta,i}=\mathbf{E}^{0}_{\beta,i}+
    \sum_{(\gamma,j)\neq(\beta, i)}
    \hat{G}(\mathbf{r}_{\beta,i},\mathbf{r}_{\gamma,j})\mathbf{P}_{\gamma,j}=\\=\mathbf{E}^{0}_{\beta,i}+
    \sum_{j\neq i}
    \hat{G}(\mathbf{r}_{\beta,i},\mathbf{r}_{\beta,j})\mathbf{P}_{\beta,j}+\\
    \sum_{\gamma\neq\beta}\sum_{j}
    \hat{G}(\mathbf{r}_{\beta,i},\mathbf{r}_{\gamma,j})\mathbf{P}_{\gamma,j},
    \label{eq:2}
\end{multline}
where $\hat{G}(\mathbf{r}_\mathrm{B},\mathbf{r}_\mathrm{A})$ is the dyadic Green’s function defining electric field induced at the point $\mathbf{r}_\mathrm{B}$ by a dipole at the coordinate $\mathbf{r}_\mathrm{A}$. Vector $\mathbf{r}_{\beta,i}$ indicates the position of the corresponding $(\beta,i)$-th particle.

Finally, in order to obtain a closed-form expression we represent the dipole moment of a particle through the background electric field acting on it and apply Bloch theorem to connect electric fields of different cells by the phase factor:

\begin{multline}
    \mathbf{E}^{\mathrm{bg}}_{\beta,i}=\mathbf{E}^{0}_{\beta,i}+
    \sum_{j\neq i}
    \hat{G}(\mathbf{r}_{\beta,i},\mathbf{r}_{\beta,j})e^{\mathrm{i}\mathbf{k}_\parallel(\mathbf{r}_{\beta,j}-\mathbf{r}_{\beta,i})}\hat{\alpha}_\beta \mathbf{E}_{\beta,i}^{\mathrm{bg}}+\\
    \sum_{\gamma\neq\beta}\sum_{j}
    \hat{G}(\mathbf{r}_{\beta,i},\mathbf{r}_{\gamma,j})e^{\mathrm{i}\mathbf{k}_\parallel(\mathbf{r}_{\gamma,j}-\mathbf{r}_{\gamma,i})}\hat{\alpha}_\gamma \mathbf{E}_{\gamma,i}^{\mathrm{bg}},
    \label{eq:3}
\end{multline}
where $\hbar \mathbf{k}_\parallel$ is an in-plane quasimomentum defined by an external wave incident on the structure. In this way, equations on the fields in different cells separate and their solutions are connected by the Bloch theorem. Therefore, we consider a system of equations for an arbitrary cell and omit its index in the equations below.
Denoting sums in an appropriate way, we express the same equation in a matrix form:
\begin{multline}\rotatebox[origin=c]{90}{\parbox[l]{1.6cm}{\centering number of particles in a cell}}
\left\{
\begin{pmatrix}
     \mathbf{E}^{\mathrm{bg}}_{\beta=1} \\ 
          \mathbf{E}^{\mathrm{bg}}_{\beta=2}\\
     \vdots
\end{pmatrix}
\right.
=
\begin{pmatrix}
     \mathbf{E}^{0}_{\beta=1} \\ 
          \mathbf{E}^{0}_{\beta=2}\\
     \vdots
\end{pmatrix}
+\\
\begin{pmatrix}
    \hat{C}_{11} & \hat{C}_{12} & \vdots \\
    \hat{C}_{21} & \hat{C}_{22} & \vdots \\
    \hdots &\hdots & \ddots
\end{pmatrix}
\begin{pmatrix}
    \hat{\alpha}_1 &0&0\\
    0 &\hat{\alpha}_2 & 0\\
    0 & 0 &\ddots
\end{pmatrix}
\begin{pmatrix}
     \mathbf{E}^{\mathrm{bg}}_{\beta=1} \\ 
          \mathbf{E}^{\mathrm{bg}}_{\beta=2}\\
     \vdots
\end{pmatrix},
    \label{eq:4}
\end{multline}
where

\begin{equation}
    \hat{C}_{\beta\gamma}(\mathbf{k}_\parallel)=
    \left\{
    \begin{array}{ll}
    \sum_{j\neq i}
    \hat{G}(\mathbf{r}_{\beta,i},\mathbf{r}_{\beta,j})e^{\mathrm{i}\mathbf{k}_\parallel(\mathbf{r}_{\beta,j}-\mathbf{r}_{\beta,i})} & \mbox{for $\beta=\gamma$}\\
    \sum_{j}
    \hat{G}(\mathbf{r}_{\beta,i},\mathbf{r}_{\gamma,j})e^{\mathrm{i}\mathbf{k}_\parallel(\mathbf{r}_{\gamma,j}-\mathbf{r}_{\gamma,i})} & \mbox{for $\beta\neq\gamma$}\\
    \end{array}
    \right.,
    \label{eq:5}
\end{equation}
Tensors $\hat{C}_{\beta\beta}$, which stand on a diagonal are the well-known dynamic interaction constants coming from the simple lattices \cite{Belov2005,fradkin2019fourier,fradkin2018fourier}. Details on possible approaches of its calculations in certain environments are discussed in our papers \cite{fradkin2019fourier,fradkin2018fourier} as well. Off-diagonal tensors $\hat{C}_{\beta\gamma}$ calculation is considered in Appendix \ref{App_A}.

Eqn. \ref{eq:4} suggests that it is natural to introduce a generalized effective polarizability, $\hat{\aleph}$, which connects dipole moments of the particles with corresponding electric fields acting on them:
\begin{equation}
\begin{pmatrix}
     \mathbf{P}_{\beta=1} \\  \mathbf{P}_{\beta=2} \\
     \vdots
\end{pmatrix}=
\hat{\aleph}
\begin{pmatrix}
     \mathbf{E}^{0}_{\beta=1} \\  \mathbf{E}^{0}_{\beta=2} \\
     \vdots
\end{pmatrix},
\end{equation}
where
\begin{multline}
\hat{\aleph}=
\begin{pmatrix}
    \hat{\alpha}_1 &0&0\\
    0 &\hat{\alpha}_2 & 0\\
    0 & 0 &\ddots
\end{pmatrix}\times\\\left[\hat{I}-
\begin{pmatrix}
    \hat{C}_{11} & \hat{C}_{12} & \vdots \\
    \hat{C}_{21} & \hat{C}_{22} & \vdots \\
    \hdots &\hdots & \ddots
\end{pmatrix}
\begin{pmatrix}
    \hat{\alpha}_1 &0&0\\
    0 &\hat{\alpha}_2 & 0\\
    0 & 0 &\ddots
\end{pmatrix}
\right]^{-1}.
\end{multline}

This quantity not only enables calculation of scattering matrix in dipole approximation but makes the qualitative analysis of main phenomena in such structures convenient as well. The physical sense of many effects can be understood from the structure of the matrices introduced above by an analogy with the paper \cite{baur2018}.

\section{Scattering matrix calculation}

The introduction of generalized effective polarizability of particle ensemble, $\aleph$, makes it easy to calculate their dipole response on the given electric field. However, our final goal is the calculation of the scattering matrix of the so-called local layer, containing lattice with local its closest dielectric environment. Here we follow up our notations and approach introduced in \cite{fradkin2019fourier,fradkin2018fourier} and report just the required modifications for the consideration of several particles in a cell.

\begin{figure}[h]
    \centering
    \includegraphics[width=1\linewidth]{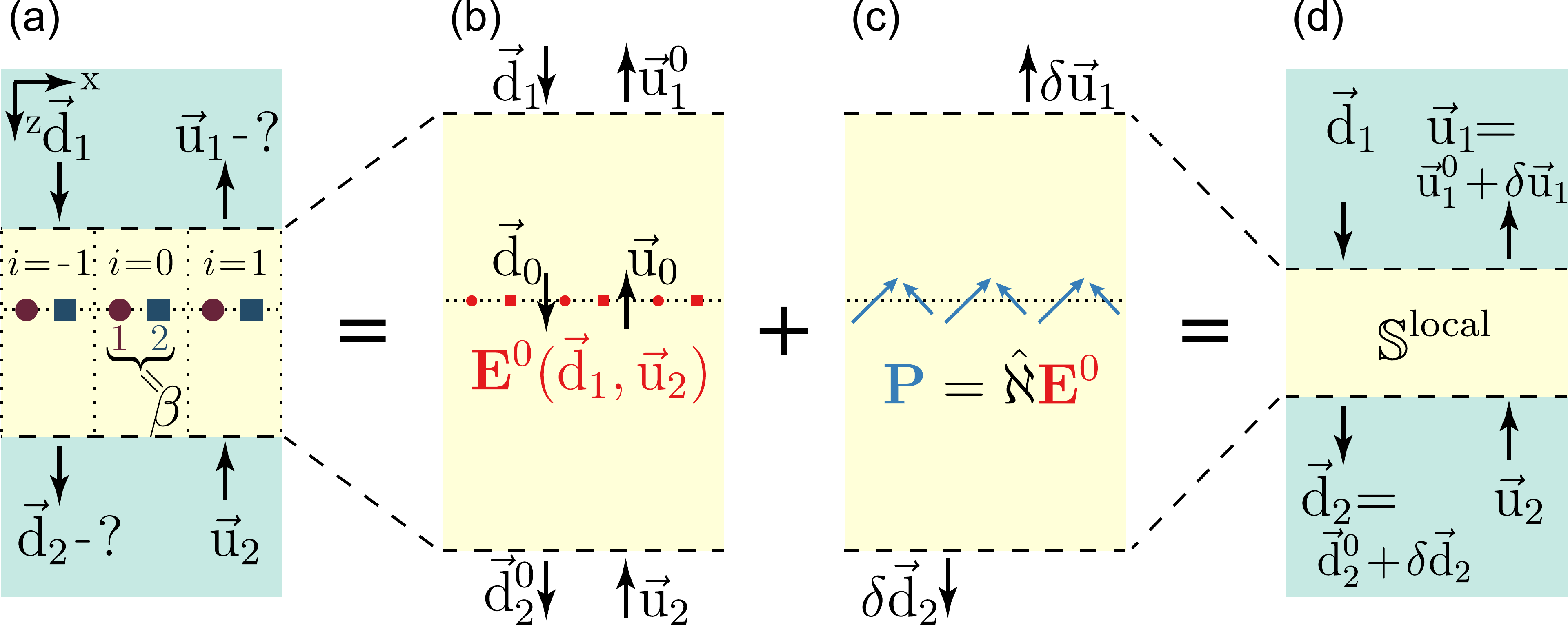}
    \caption{Calculation of the scattering matrix of a plasmonic lattice in dipole approximation. (a) Nanoparticles lattice in a layered medium. (b) Calculation of the external field at the position of nanoparticles (red points) in the layered medium without nanoparticles. (c) Calculation of dipole moments of nanoparticles (blue arrows). (d) Calculation of the local scattering matrix $\mathbb{S}^{\mathrm{local}}$. In panels (a)-(d) dashed lines separate the local dielectric environment of the nanoparticles (yellow color) from the outer dielectric environment (green color). Both local and outer dielectric environments might include any number of vertically homogeneous layers and interfaces between them.}
    \label{fig:sketch}
\end{figure}

The basic idea reported in \cite{fradkin2019fourier,fradkin2018fourier} is that we need to artificially insulate the local layer of the lattice from the whole structure (see Fig. \ref{fig:sketch}). This layer should include the lattice itself as well as its closely-disposed ambiance (such as an interface on which it lays) in order to get rid of high-$\mathbf{k}_\parallel$ harmonics on its boundaries. In other words, the scattering matrix of this layer will have a very limited number of elements corresponding to low-$\mathbf{k}_\parallel$ harmonics that can not be neglected. This matrix can be represented as a sum of two terms - the first one is a scattering matrix of the background layer, which corresponds to a structure in the absence of the lattice and the second one is obviously an addition from this lattice:

\begin{equation}
    \mathbb{S}^{\mathrm{local}}=\mathbb{S}^{\mathrm{local}}_0+\delta\mathbb{S}^{\mathrm{local}}.
\end{equation}
In turn, this addition can be found as:
\begin{equation}
    \delta\mathbb{S}^{\mathrm{local}}=\mathbb{B}^{\mathrm{out}}\mathbb{A}\mathbb{B}^{\mathrm{in}},
\end{equation}
where $\mathbb{B}^{\mathrm{in}}$ helps to obtain the vector of amplitudes of up- and down-going waves at the plane of a lattice in the background layer from amplitudes of the incoming wave at the boundaries of this layer; the matrix $\mathbb{A}$ transfers this amplitudes to discontinuities induced by currents in the lattice itself and finally $\mathbb{B}^{\mathrm{out}}$ binds these discontinuities with amplitudes of outgoing waves on the boundaries of the local layer. The wonderful fact is that expression for $\mathbb{B}^{\mathrm{out}}$ and $\mathbb{B}^{\mathrm{in}}$ are exactly the same as in \cite{fradkin2019fourier,fradkin2018fourier} for simple lattices and depend only on the structure of the background layer. $\mathbb{A}$ can be represented as  $\hat{A}$ tensor sandwiched between the material matrix of the background layer, $\mathbb{F}$, and its inverse one:
\begin{equation}
    \mathbb{B}=\mathbb{F}^{-1}\hat{A}\mathbb{F},
\end{equation}
where $\hat{A}$ is an analogous of $\mathbb{A}$ operating not with waves' amplitudes but with field harmonics. In this way, only the matrix $\hat{A}$, which contains all the information about the lattice itself, is changed when a lattice contains more than one particle. However, its structure remains the same as for simple lattices in the original paper \cite{fradkin2019fourier,fradkin2018fourier}. It is easy to understand that in our case $\hat{A}$ tensor has the following form:

\begin{multline}
    \hat{A} = \frac{-4 \pi \mathrm{i}  k_0}{s}\begin{bmatrix}
    \hat{0} & \hat{0}& \hat{K}_x/\varepsilon\\
    \hat{0} & \hat{0}& \hat{K}_y/\varepsilon\\
    \hat{0} & \hat{I}&\hat{0} \\
    {-}\hat{I} & \hat{0} &\hat{0} 
    \end{bmatrix}
    \begin{bmatrix}
    \vec{\Gamma}_1 & \vec{0} & \vec{0}\\
    \vec{0} & \vec{\Gamma}_1 & \vec{0}\\
    \vec{0} & \vec{0} & \vec{\Gamma}_1\\
    \vec{\Gamma}_2 & \vec{0} & \vec{0}\\
    \vec{0} & \vec{\Gamma}_2 & \vec{0}\\
    \vec{0} & \vec{0} & \vec{\Gamma}_2\\
    \hdotsfor{3}\\
    \end{bmatrix}^{\dagger}
    \hat{\aleph}\times \\
    \begin{bmatrix}
    \vec{\Gamma}_1 & \vec{0} & \vec{0}\\
    \vec{0} & \vec{\Gamma}_1 & \vec{0}\\
    \vec{0} & \vec{0} & \vec{\Gamma}_1\\
    \vec{\Gamma}_2 & \vec{0} & \vec{0}\\
    \vec{0} & \vec{\Gamma}_2 & \vec{0}\\
    \vec{0} & \vec{0} & \vec{\Gamma}_2\\
    \hdotsfor{3}\\
    \end{bmatrix}
    \begin{bmatrix}
    \hat{I} & \hat{0} & \hat{0}  & \hat{0} \\
    \hat{0} & \hat{I} & \hat{0}  & \hat{0} \\
    \hat{0} & \hat{0} & \hat{K}_y/\varepsilon & -\hat{K}_x/\varepsilon \\
    \end{bmatrix},
\end{multline}
where $s$ is an area of a unit cell in a real space, $k_0$ is a wave vector in vacuum, $\varepsilon$ is permittivity of the homogeneous medium in which the lattice is located (upper medium for the case of a lattice on an interface \cite{fradkin2019fourier,fradkin2018fourier}), dagger denotes Hermitian conjugate. $K_x$ and $K_y$ are dimensionless diagonal operators \cite{tikhodeev2002}:
\begin{equation}
    \hat{K}_x = \frac{1}{k_0} \mathrm{diag}(k_y +\vec{g}_x ),    \hat{K}_y = \frac{1}{k_0} \mathrm{diag}(k_y +\vec{g}_y ),
\end{equation}
where $\vec{g}_x$ and $\vec{g}_y$ are $1 \times N_g$ hypervectors of $x$ and $y$ projections of reciprocal lattice vectors. $\vec{\Gamma}_\beta$ calculates electric field at the position of the $\beta$-th particle from the Fourier harmonics and is set by the following expression:
\begin{equation}
    \vec{\Gamma}_\beta = \mathrm{exp}(i(k_x+\vec{g}_x)x_\beta+i(k_y+\vec{g}_y)y_\beta),
\end{equation}
where $(x_\beta,y_\beta)$ are the coordinates of the $\beta$ particle inside the unit cell in real space. We omit the second index numbering the cells since a particle in any cell can be considered with the same result.

Referring the reader to \cite{fradkin2019fourier,fradkin2018fourier} for the detailed description of this matrix derivation, here we just describe the purpose of each multiplier from right to left. At first, amplitudes of harmonics of $x$ and $y$ components of electric and magnetic fields are transformed to harmonics of all the three components of electric fields. After that, we apply the inverse Fourier transform to determine the electric field at the position of the particles in a cell and obtain the dipole moment by applying already discussed $\aleph$. After that, in reverse order, we calculate the Fourier harmonics of dipole moments and transfer them to discontinuities in electric and magnetic fields that they induce \cite{lobanov2012emission,taniyama2008s}.

\section{Example: routing  plasmonic metasurface}

An ability to consider plasmonic lattices in dipole approximation allows us
to observe a large number of exciting physical phenomena. This flexibility is provided primarily by the variation of the particle's shape from symmetric discs to strongly prolongated nanobars, their orientation and relative position of sublattices. In this way, the superposition of the optical properties of individual nanoparticle and the geometry of a lattice determines the behavior of guided modes excitation and in turn spectra of the photonic crystals.

\begin{figure}
    \centering
    \includegraphics[width=\linewidth]{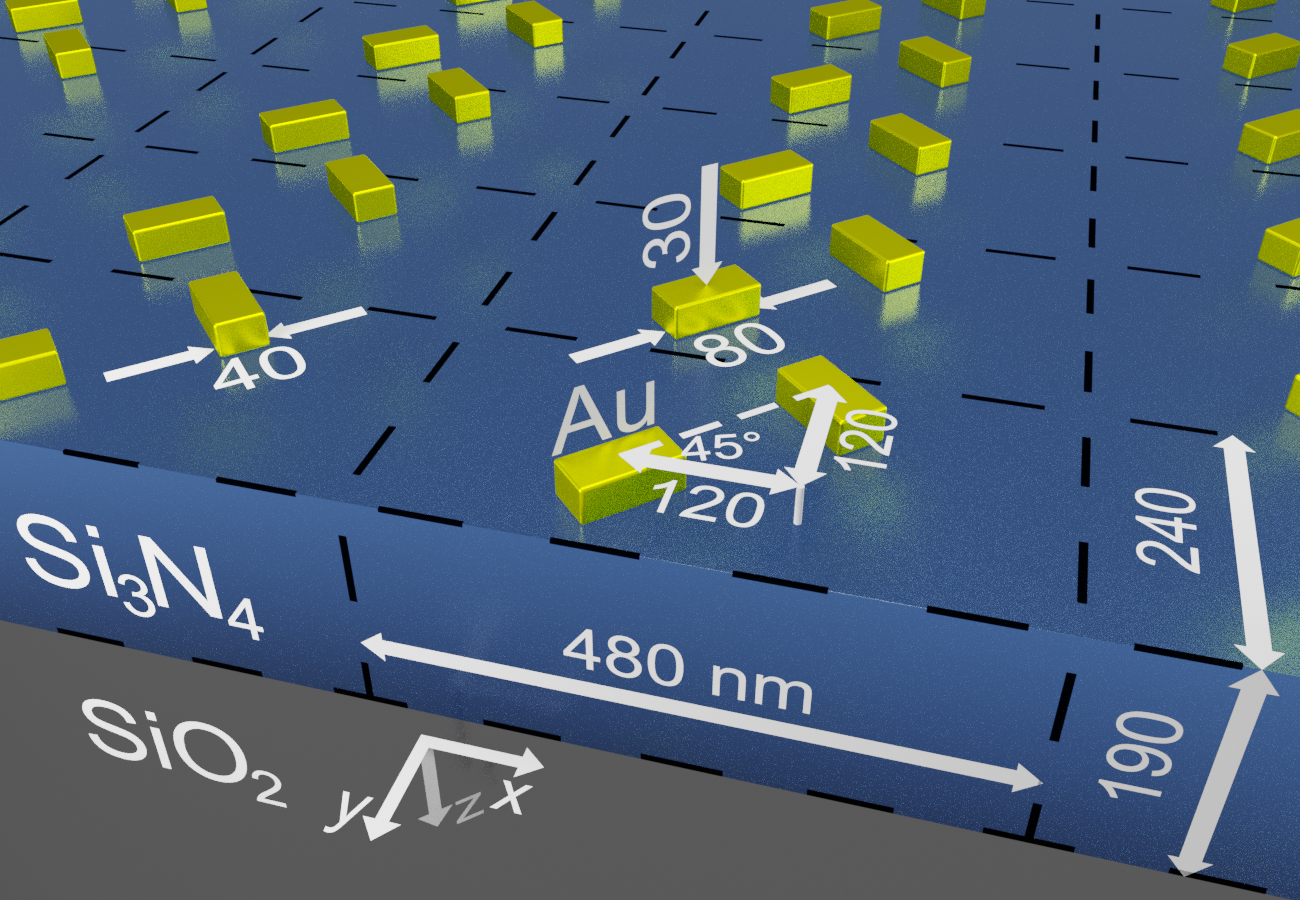}
    \caption{Schematic of the plasmonic structure with a basis on a waveguide. All the proportions are shown in the picture in nanometers. Edges of plasmonic nanobars are smoothed out with 1 nm fillet.}
    \label{fig:scheme}
\end{figure}

\begin{figure*}[t!]
    \centering
    \includegraphics[width=\textwidth]{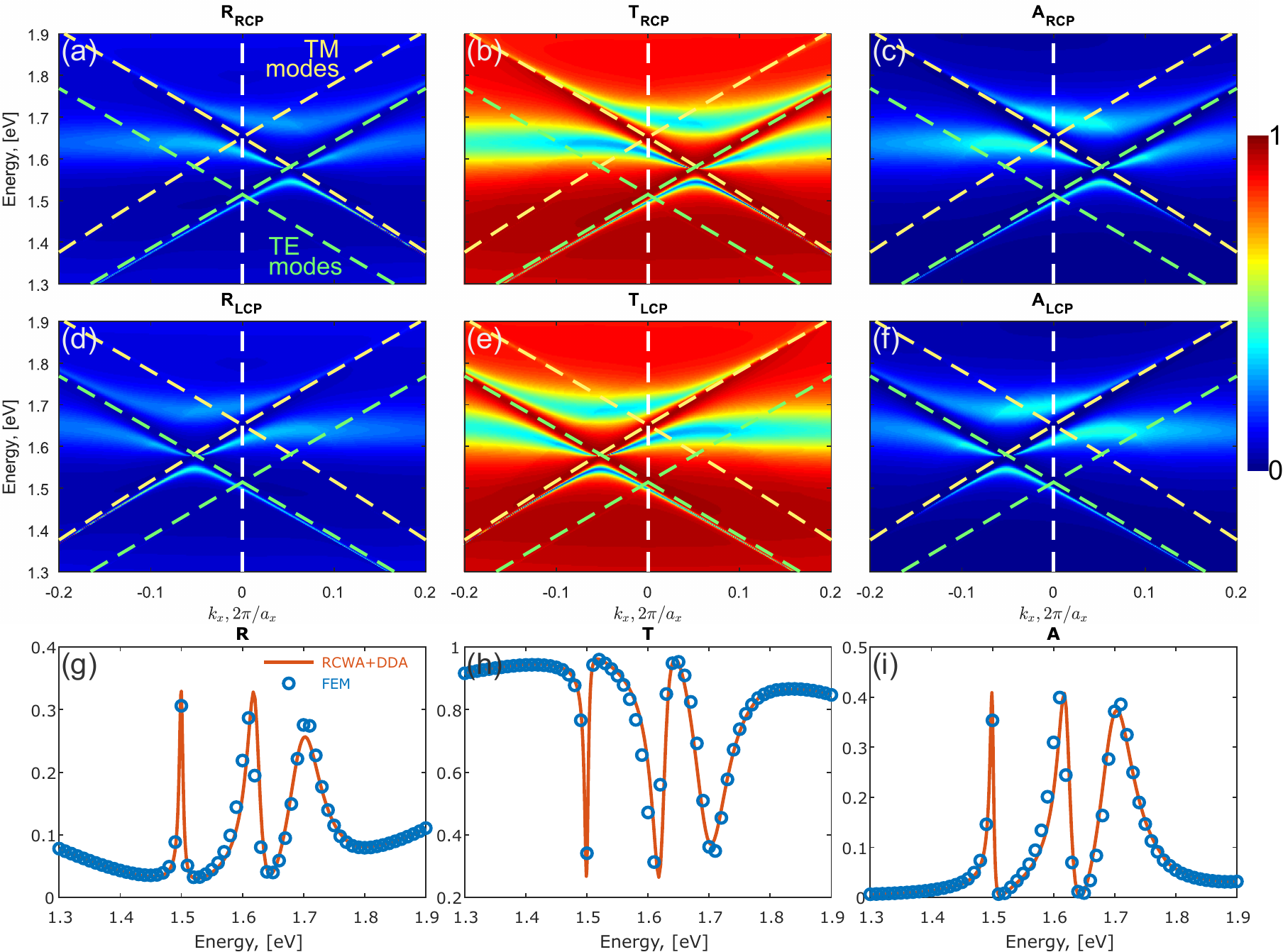}
    \caption{In-plane wave-vector and energy dependencies of reflection (a,d), transmission (b,e) [in the main diffraction order] and absorption (c,f) in right-handed (a-c) and left-handed (d-f) circular polarizations for the plasmonic structure on a waveguide. Green and yellow dashed lines correspond to TE and TM modes of a waveguide without plasmonic lattice transferred to the first Brillouin zone.
    The color scale of (a)–(f) is explained on the right. (g)–(i) show spectra for the normal incidence of light, which corresponds to $k_x=0$ section of angle-dependent spectra. They include the comparison of computations conducted via our approach with conventional FEM calculations in COMSOL Multiphysics.}
    \label{fig:2}
\end{figure*}

Here, we demonstrate the performance of our approach on a lattice of two nanobars, which supports the effect often referred to as photonic spin-orbit interaction  \cite{bliokh2015spin,shitrit2013spin}. Our lattice is placed onto a $190 \mathrm{nm}$-height $\mathrm{Si}_3\mathrm{N}_4$ ($\varepsilon_{\mathrm{Si}_3\mathrm{N}_4}=4.1$) waveguide on a silica ($\varepsilon_{\mathrm{Si}\mathrm{O}_2}=2.1$) substrate. Gold nanobars of $80\times40\times30 \mathrm{nm}^3$ sizes are positioned according to the Fig. \ref{fig:scheme} and are described by Johnson-Christy optical constants \cite{JohnsonChristy1972}. In order to avoid an appearance of strongly localized mesh-sensitive edge plasmons and simplify FEM calculation of nanobar's polarizability tensor and reference spectra (see Fig. \ref{fig:2} (g-i)), which will be considered below, we smoothed out its edges with 1 nm fillets.

The general concept of an effect was presented in the pioneering paper \cite{lin2013polarization}, where the authors proposed the 2D periodic structure consisting of two sublattices of perpendicular slots in metal. Because of their perpendicularity, normally incident circularly polarized light is scattered with a $\pi/2$ phase shift by slots that belong to different sublattices. At the same time relative shift of a quarter of the period along the $x$-axis (same as in Fig. \ref{fig:scheme}) results in an additional $\pi/2$ phase shift in the excitation of guided modes (surface plasmon polariton modes in the original paper) propagating along and against the $x$-axis. Finally,
this means, that sublattices excite guided modes that interfere constructively in one direction (in which phase shifts compensate each other) and destructively in the opposite one (in which phases add up and result in $\pi$ shift). In other words, this structure routes light of each circular polarization to corresponding guided modes propagating in opposite directions.

Following the original idea, we illuminate the structure from above by light of both circular polarizations and observe reflection, transmission in the main diffraction order and absorption spectra (see Fig. \ref{fig:2} (a-f)). Panels (g-i) show the comparison of spectra obtained by our approach with reference points calculated by FEM in COMSOL Multiphysics. We see that the match is very good across the whole spectrum and even the narrow lines obtained by two approaches coincide. Moreover, since FEM is a numerical method as well and has its own finite precision, we even do not know which one is more precise. The difference between the two approaches can be considered as a rough estimation of an error. However, in a case when each of the alternative approaches provides precision, which is sufficient to observe the characteristic phenomena an issue of computation time becomes the most essential one. FEM calculations require high computational power and consume a lot of RAM. That is why it takes about a minute to compute one point on a high-power PC. Our approach requires precomputations of polarizability tensor, which take from several minutes to an hour in COMSOL Multiphysics depending on the required accuracy and available computational resources. However, after that, it becomes possible to obtain the 6 spectra like in Fig. \ref{fig:2} (a-f) each consisting of $500\times250=150000$ points for any lattice from the considered particles in approximately 10 minutes on a regular laptop. Such pictures as Fig. \ref{fig:2} (g-i) each consisting of 300 points are plotted in approximately 1 minute.

It can be seen from Fig. \ref{fig:2} (a-f) that for circular polarization of an incident light $k_x$-dependent spectra do not have mirror symmetry with respect to $k_x=0$ plane. Moreover, in accordance with the symmetry of the structure all the spectra for left-hand polarized light (Fig. \ref{fig:2} (d-f)) are mirror images of corresponding right-hand circular polarization spectra (Fig. \ref{fig:2} (a-c)).
This phenomenon is clearly seen from the dispersion curves of the quasiguided modes. Indeed, the TE mode propagating to the right and the TM mode propagating to the left are excited by right-hand polarized light and vice versa for left-hand one. Thus, we observe the desired phenomenon of spin-orbit coupling. However, in contrast to many works on surface plasmon polaritons \cite{lin2013polarization, lee2015plasmonic,zhao2019compound,mueller2014polarization,yin2018polarization,miroshnichenko2013polarization,zhao2019polarization,huang2013helicity,ding2018bifunctional} we have two modes of different polarizations that behave differently as well. In order to explain the considered phenomena, we build a toy model in Appendix \ref{App_B}. Another exciting phenomenon is the strong coupling of localized plasmonic resonance, TE and TM modes, which is observed on all the spectra. Most probably dipole approximation provides us with a well-known bound state in the continuum (BIC) \cite{hsu2016bound} in this case. However, we leave consideration of these high-quality modes for further research.

Besides the fact that such a structure supports so many interesting physical phenomena, they can be implemented for practical purposes. First of all, this lattice can be used as a grating coupler for normally incident light in contrast to conventional ones, which work mostly with oblique incidence. Secondly, the direction of the coupling is controlled by the polarization of light, which can be used as a convenient tool for control of light propagation. Moreover, the excitation of TE and TM modes in opposite directions by a given circular polarization makes the toolset even wider. This one single example demonstrates the high diversity of possible applications that can be considered by the means of the proposed approach.

\section*{Conclusion}

In this paper, we have applied the calculation method combined of DDA and FMM for the consideration of periodic photonic structures with inclusions of dipole nanoparticles lattice with basis. This approach allows obtaining spectra with extremely high energy and angular resolution as it is much faster than conventional methods.
Direct comparison of spectra with FEM-calculated ones demonstrated the high precision of calculations, which makes the method reliable and applicable for practical utilization. In particular, our approach is very useful for plasmonic lattices with bases, which was shown by a consideration of a lattice supporting the photonic spin-orbit coupling. This structure can be used as a circular polarization splitter for normally incident light or as a grating coupler routing TE and TM modes in opposite directions. It shows the potential of our approach for the design of practical photonic devices.

\section*{Acknowledgements}
This work was supported by the Russian Foundation for Basic Research (Grant No. 18-29-20032).

\appendix

\section{Calculation of generalized dynamic interaction constant}
\label{App_A}

The concept of generalized dynamic interaction constant for the lattice with basis is very important. However, for its practical implementation in computational approaches, we need an efficient method for its calculation. Each block of the whole tensor arising in Eqn. \ref{eq:4} has the form defined by Eqn. \ref{eq:5}.

Since we consider dipoles lattices located only in environments that are translation invariant in $x-y$ plane then the Green’s function can be taken to be a convolution kernel. This leads to an equality of all the diagonal blocks, which correspond to the self-action of a certain sublattice since they do not depend on $\beta$:
\begin{multline}
    \hat{C}_{\beta\beta}    =\sum_{j\neq i}
    \hat{G}(\mathbf{r}_{\beta,i},\mathbf{r}_{\beta,j})e^{\mathrm{i}\mathbf{k}_\parallel(\mathbf{r}_{\beta,j}-\mathbf{r}_{\beta,i})}=\\
    \sum_{j\neq i}
    \hat{G}(\mathbf{r}_{\beta,i}-\mathbf{r}_{\beta,j})e^{\mathrm{i}\mathbf{k}_\parallel(\mathbf{r}_{\beta,j}-\mathbf{r}_{\beta,i})}=\\
    \sum_{j\neq i}
    \hat{G}(\mathbf{t}_{i-j})e^{-\mathrm{i}\mathbf{k}_\parallel(\mathbf{t}_{i-j})}= \\   \sum_{j\neq 0}
    \hat{G}(\mathbf{t}_{j})e^{-\mathrm{i}\mathbf{k}_\parallel(\mathbf{t}_{j})},
\end{multline}
where $\mathbf{t}_j$ is the $j$-th translational vector of the lattice.
Their calculation was already discussed in many papers and one of the approaches was considered in our work \cite{fradkin2019fourier,fradkin2018fourier}. Calculation of off-diagonal blocks is much easier since summation is held over the whole lattice without exclusions:

\begin{multline}    
    \hat{C}_{\beta\gamma}(\mathbf{k}_\parallel)=\sum_{j}
    \hat{G}(\mathbf{r}_{\beta,i},\mathbf{r}_{\gamma,j})e^{\mathrm{i}\mathbf{k}_\parallel(\mathbf{r}_{\gamma,j}-\mathbf{r}_{\gamma,i})}=\\ \sum_j \hat{G}(\mathbf{r}_{\beta,i}-\mathbf{r}_{\gamma,j})e^{\mathrm{i}\mathbf{k}_\parallel(\mathbf{r}_{\gamma,j}-\mathbf{r}_{\gamma,i})}=\\
    \sum_j \hat{G}(\mathbf{r}_{\beta}-\mathbf{r}_{\gamma}+\mathbf{t}_{j})e^{-\mathrm{i}\mathbf{k}_\parallel\mathbf{t}_{j}}.
    \label{eq:App_A_1}
\end{multline}

In principle, the sum above can be calculated as is, however, in real space Green's function decays very slowly and therefore convergence rate will be poor. The classical approach to cope with this problem in a homogeneous environment was formulated by Ewald \cite{ewald1921} and implemented in many papers \cite{Belov2005,stevanovic2006,papanicolaou1999,wette1965}. Here, we decided to apply another, very easy-to-implement method, which is applicable for a lattice placed onto an interface.

So, the sum in \ref{eq:App_A_1} can be expressed as a sum over reciprocal lattice via the Poisson formula:
\begin{equation}    
    \hat{C}_{\beta\gamma}(\mathbf{k}_\parallel)=\sum_{j}\frac{4\pi^2}{s}\hat{M}(\mathbf{k}_\parallel+\mathbf{g}_j)e^{\mathrm{i} (\mathbf{k}_\parallel+\mathbf{g}_j) (\mathbf{r}_\beta-\mathbf{r}_\gamma)},
\end{equation}
where $\mathbf{g}_j$ is $j$-th vector of reciprocal lattice and $\hat{M}(\mathbf{k}_\parallel)=\frac{1}{4\pi^2}\int \hat{G}(\mathbf{r_\parallel})e^{-\mathrm{i}\mathbf{k}_\parallel\mathbf{r}_\parallel}d^2\mathbf{r}_\parallel$ is dyadic Green's function in reciprocal space. Nevertheless, the new sum in reciprocal space converges slowly as well and some components of the tensor do not converge at all. This is due to the fact that technically dyadic Green's function is not $L^2$-integrated function and therefore does not have a Fourier image from a mathematical point of view.
However, high-$k_\parallel$ harmonics are mainly associated with a local high-gradient field in the vicinity of point dipoles. Therefore, if we exclude them by smooth low-spatial-frequency filter $F(k_\parallel)$ then the interaction between particles will remain almost unchanged, whereas asymptotic convergence rate can be made as fast as needed:

\begin{equation}    
    \hat{C}_{\beta\gamma}(\mathbf{k}_\parallel)\approx\sum_{j}\frac{4\pi^2}{s}F(|\mathbf{k}_\parallel+\mathbf{g}_j|)\hat{M}(\mathbf{k}_\parallel+\mathbf{g}_j)e^{\mathrm{i} (\mathbf{k}_\parallel+\mathbf{g}_j) (\mathbf{r}_\beta-\mathbf{r}_\gamma)}.
    \label{eq:App_A_2}
\end{equation}

This formula can be used for practical calculations, but how does this filtering worsen the result? To figure it out, we need to return to real space. Indeed, multiplication in Fourier space is equivalent to a convolution in real space. This means that we substitute dipole moment density defined by delta function $\delta(\mathbf{r})$ with a source distributed in accordance with inverse Fourier image of filtering function $f(\mathbf{r}) = \int F(\mathbf{
k}_\parallel)e^{\mathrm{i} \mathbf{k}_\parallel\mathbf{r}_\parallel}d^2\mathbf{k}_\parallel$.

On the one hand, it is obvious that the narrower is $f(\mathbf{r}_\parallel)$ the preciser are results, but on the other hand, we understand that in reality the dipole moment is not located at the point as well and therefore there is no sense to make the distribution area of a source much smaller than the size of a particle. At the same time, the wider is a filter in real space the narrower it is in reciprocal one and therefore the faster is convergence.
Calculation of \ref{eq:App_A_2} is relatively fast operation since all the summands are analytical functions, it does not include integrals or other coast operations and as a result does not limit the speed of calculations.
Therefore, we recommend choosing relatively narrow filter in real space. The reasonable half-width of $f(\mathbf{r}_\parallel)$ for wide range of problems is $10-20\mathrm{nm}$.

The shape of the filter can be arbitrary but has to meet certain requirements. In reciprocal space $F(\mathbf{k}_\parallel)$ should be equal to unity for small wavevectors and decay to zero for high ones. It results in a fixed integral $\int f(\mathbf{r}_\parallel) d^2r=4\pi^2$  in real space, which corresponds to the preservation of the total dipole moment ($4\pi^2$ is determined by the choice of the constant in Fourier transform). Also, the width of the real-space filter should be much smaller than the distance to the nearest neighbor and wavelength.

For our implementation we have chosen the Gaussian filter since it provides fast decay in both real and reciprocal spaces:
\begin{equation}
    f(\mathbf{r}_\parallel)=\frac{2\pi}{\sigma^2}e^{-\frac{r^2_\parallel}{2\sigma^2}}, \qquad F(\mathbf{k}_\parallel)=e^{-\frac{\sigma^2 k^2_\parallel}{2}},
\end{equation}
where $\sigma$ is a typical half-width of the blurred dipole source in real space. Thus, off-diagonal  blocks $\hat{C}_{\beta \gamma}$ can be found as follows:

\begin{equation}    
    \hat{C}_{\beta\gamma}(\mathbf{k}_\parallel)\approx\sum_{j}\frac{4\pi^2}{s}
    e^{-\frac{\sigma^2 (\mathbf{k}_\parallel+\mathbf{g}_j)^2}{2}}
    \hat{M}(\mathbf{k}_\parallel+\mathbf{g}_j)e^{\mathrm{i} (\mathbf{k}_\parallel+\mathbf{g}_j) (\mathbf{r}_\beta-\mathbf{r}_\gamma)}.
\end{equation}

\section{Dipole toy model}
\label{App_B}

In this appendix, we develop a toy model based on dipole approximation to explain qualitatively the observed effects in spectra from Fig. \ref{fig:2}. We consider a system of equations on dipole moments analogous to Eqns. \ref{eq:1}-\ref{eq:2}, but for the whole structure including the waveguide. For practical calculations, there is no sense to include waveguide into consideration, at least because of the necessity to fix its height and material of the substrate, which can be easily varied within FMM. Also, since the bottom of the waveguide is far from the lattice, it's inclusion into consideration will not decrease the number of harmonics nor increase the precision. However, here we are not going to repeat the procedure outlined in this paper, but just to focus on the main peculiarities of the obtained equations and their solutions. In this way, dipole moments of the particles are determined as a response on a sum of the electric field induced by external light and field rescattered by the lattice:

\begin{equation}
    \begin{bmatrix}
         \mathbf{P}_1\\
         \mathbf{P}_2
    \end{bmatrix} = 
    \begin{bmatrix}
        \hat{\alpha}^{\mathrm{wg}}_1 &0\\
        0 & \hat{\alpha}^{\mathrm{wg}}_2
    \end{bmatrix}
    \left(    \begin{bmatrix}
         \mathbf{E}^0_1\\
         \mathbf{E}^0_2
    \end{bmatrix}+
    \begin{bmatrix}
         \mathbf{E}^{\mathrm{lattice}}_1\\
         \mathbf{E}^{\mathrm{lattice}}_2
    \end{bmatrix}
        \right),
        \label{eq:App_B_1}
\end{equation}
where indices 1 and 2 correspond to particles positioned at $(0,0)$ and $(a_x/4,-a_y/2)$ coordinates correspondingly (see. Fig. \ref{fig:scheme}). Polarizabilities of particles $\hat{\alpha}^{\mathrm{wg}}_{1,2}$ should be calculated for the case when they are placed on a waveguide. However, since the bottom of the waveguide is far from the particles we can assume that $\hat{\alpha}^{\mathrm{wg}}_{1,2}=\hat{\alpha}_{1,2}$ with a good precision, where the latter polarizabilities correspond to particles on an interface between two half-spaces of different media and were used in calculations in the article. Moreover, from the numerical values of these tensors we know, that for the sake of simplicity it is possible to consider particles as uniaxial scatterers:
\begin{equation}
    \hat{\alpha}_1 = 
    \frac{\alpha}{2}
    \begin{bmatrix}
    1 &-1\\
    -1 & 1
    \end{bmatrix}, \quad
    \hat{\alpha}_2 = 
    \frac{\alpha}{2}
    \begin{bmatrix}
    1 & 1\\
    1 & 1
    \end{bmatrix},
    \label{eq:App_B_2}
\end{equation}
where $\alpha$ is their polarizability along the $80\mathrm{nm}$-long side.

Here, we consider illumination of the structure by normally incident light with energy close to either TE or TM mode excitation, which corresponds to intersections of white dotted line with green and yellow ones in the Fig. \ref{fig:2}. 
Under the resonant condition the contribution of the lattice into the background electric field acting on the particles in the considered cell can be expressed as a sum of fields of two modes propagating in opposite directions:

\begin{equation}
    \begin{bmatrix}
         \mathbf{E}^{\mathrm{lattice}}_1\\
         \mathbf{E}^{\mathrm{lattice}}_2
    \end{bmatrix} = 
    A^\rightarrow \begin{bmatrix}
         \Tilde{\mathbf{E}}^\rightarrow_1\\
         \Tilde{\mathbf{E}}^\rightarrow_2
    \end{bmatrix}+
    A^\leftarrow \begin{bmatrix}
         \Tilde{\mathbf{E}}^\leftarrow_1\\
         \Tilde{\mathbf{E}}^\leftarrow_2
    \end{bmatrix}.
    \label{eq:App_B_3}
\end{equation}

According to the theory of quasinormal modes \cite{lalanne2018,sauvan2013, Bai2013}  and especially their application for the case of Fano resonances in periodic structures \cite{ weiss2016dark,weiss2017analytical}
amplitude of each mode is proportional to the product of the particles' dipole moments and field of a conjugate mode (propagating in the opposite direction):
\begin{equation}
    \begin{bmatrix}
        A^\rightarrow\\
        A^\leftarrow
    \end{bmatrix}
    = \frac{1}{\omega-\Tilde{\omega}}\frac{1}{N}
    \begin{bmatrix}
    \Tilde{\mathbf{E}}^\leftarrow_1 & \Tilde{\mathbf{E}}^\rightarrow_1\\
    \Tilde{\mathbf{E}}^\leftarrow_2 & \Tilde{\mathbf{E}}^\rightarrow_2
    \end{bmatrix}^{\mathrm{T}}
    \begin{bmatrix}
    \mathbf{P}_1\\ \mathbf{P}_2
    \end{bmatrix},
    \label{eq:App_B_4}
\end{equation}
where $\Tilde{\omega}$ is the real-valued frequency of the resonant modes and $N$ is the normalizing factor. Substituting Eqns. \ref{eq:App_B_1} and \ref{eq:App_B_3} in the latter equations we obtain the following system of equations:

\begin{multline}
    \begin{bmatrix}
    A^\rightarrow\\
    A^\leftarrow
    \end{bmatrix} = \frac{1}{\omega-\Tilde{\omega}}\frac{1}{N}
    \begin{bmatrix}
    \Tilde{\mathbf{E}}^\leftarrow_1 & \Tilde{\mathbf{E}}^\rightarrow_1\\
    \Tilde{\mathbf{E}}^\leftarrow_2 & \Tilde{\mathbf{E}}^\rightarrow_2
    \end{bmatrix}^{\mathrm{T}}
    \begin{bmatrix}
        \hat{\alpha}_1 &0\\
        0 & \hat{\alpha}_2
    \end{bmatrix}
    \begin{bmatrix}
         \mathbf{E}^0_1\\
         \mathbf{E}^0_2
    \end{bmatrix}
    +\\
\frac{1}{\omega-\Tilde{\omega}}\frac{1}{N}     \begin{bmatrix}
    \Tilde{\mathbf{E}}^\leftarrow_1 & \Tilde{\mathbf{E}}^\rightarrow_1\\
    \Tilde{\mathbf{E}}^\leftarrow_2 & \Tilde{\mathbf{E}}^\rightarrow_2
    \end{bmatrix}^{\mathrm{T}}
    \begin{bmatrix}
        \hat{\alpha}_1 &0\\
        0 & \hat{\alpha}_2
    \end{bmatrix}
        \begin{bmatrix}
    \Tilde{\mathbf{E}}^{\rightarrow }_1 & \Tilde{\mathbf{E}}^{\leftarrow }_1\\
    \Tilde{\mathbf{E}}^{\rightarrow }_2 & \Tilde{\mathbf{E}}^{\leftarrow }_2
    \end{bmatrix}\begin{bmatrix}
    A^\rightarrow\\
    A^\leftarrow
    \end{bmatrix}.
\end{multline}
And easily solve it:

\begin{multline}
    \begin{bmatrix}
    A^\rightarrow\\
    A^\leftarrow
    \end{bmatrix} =\\
    \left (
    (\omega-\Tilde{\omega})N
    \begin{bmatrix}
        1 &0\\
        0 & 1
    \end{bmatrix}-  
    \begin{bmatrix}
    \Tilde{\mathbf{E}}^\leftarrow_1 & \Tilde{\mathbf{E}}^\rightarrow_1\\
    \Tilde{\mathbf{E}}^\leftarrow_2 & \Tilde{\mathbf{E}}^\rightarrow_2
    \end{bmatrix}^{\mathrm{T}}
    \begin{bmatrix}
        \hat{\alpha}_1 &0\\
        0 & \hat{\alpha}_2
    \end{bmatrix}
        \begin{bmatrix}
    \Tilde{\mathbf{E}}^{\rightarrow }_1 & \Tilde{\mathbf{E}}^{\leftarrow }_1\\
    \Tilde{\mathbf{E}}^{\rightarrow }_2 & \Tilde{\mathbf{E}}^{\leftarrow }_2
    \end{bmatrix}
    \right)^{-1}\times\\
    \begin{bmatrix}
    \Tilde{\mathbf{E}}^\leftarrow_1 & \Tilde{\mathbf{E}}^\rightarrow_1\\
    \Tilde{\mathbf{E}}^\leftarrow_2 & \Tilde{\mathbf{E}}^\rightarrow_2
    \end{bmatrix}^{\mathrm{T}}
    \begin{bmatrix}
        \hat{\alpha}_1 &0\\
        0 & \hat{\alpha}_2
    \end{bmatrix}
    \begin{bmatrix}
         \mathbf{E}^0_1\\
         \mathbf{E}^0_2
    \end{bmatrix}.
\end{multline}
The last what left to do is to substitute the explicit expressions for polarizabilities and the modes' fields. The case of TE modes excitation gives us:
\begin{equation}
    \Tilde{\mathbf{E}}^{\mathrm{TE},\rightleftarrows}_1=
    \begin{bmatrix}
    0\\1
    \end{bmatrix}, \qquad 
    \Tilde{\mathbf{E}}^{\mathrm{TE},\rightleftarrows}_2=
    \begin{bmatrix}
    0\\e^{\pm \mathrm{i} \frac{2\pi}{a_x}\frac{a_x}{4}}
    \end{bmatrix}=
    \begin{bmatrix}
    0\\\pm \mathrm{i}
    \end{bmatrix},
\end{equation}
which results in:
\begin{equation}
\begin{bmatrix}
A^{\mathrm{TE},\rightarrow}\\
A^{\mathrm{TE},\leftarrow}
\end{bmatrix}=
    \frac{1}{\omega-\Tilde{\omega}-\alpha/N^{\mathrm{TE}}}\frac{1}{N^{\mathrm{TE}}}
    \begin{bmatrix}-1 & 1  &-\mathrm{i} & -\mathrm{i}\\
    -1 & 1  &\mathrm{i} & \mathrm{i}
    \end{bmatrix}    \begin{bmatrix}
         \mathbf{E}^0_1\\
         \mathbf{E}^0_2
    \end{bmatrix}.
\end{equation}
The resonant denominator $\omega-\Tilde{\omega}-\alpha/N$ corresponds to the hybridized lattice-waveguide resonance. The frequency of hybridized resonance is equal to $\Tilde{\omega}_{\mathrm{HR}}=\Tilde{\omega}+\alpha/N$ and is complex because of the losses on leakage and dissipation. Since polarizabilities of particles are relatively small this frequency is typically rather close to the frequency of the original mode of the waveguide.

Given that RCP light normally incident on the structure induces external field $\mathbf{E}^{0,\mathrm{RCP}}_1=\mathbf{E}^{0,\mathrm{RCP}}_2\propto \frac{\sqrt{2}}{2}\begin{bmatrix}1 \\-\mathrm{i} \end{bmatrix}$, we obtain:

\begin{equation}
\begin{bmatrix}
A^{\mathrm{TE},\rightarrow}\\
A^{\mathrm{TE},\leftarrow}
\end{bmatrix}\propto
    \frac{-(1+\mathrm{i})\sqrt{2}}{\omega-\Tilde{\omega}-\alpha/N^{\mathrm{TE}}}\frac{1}{N^{\mathrm{TE}}}\begin{bmatrix} 1\\
     0
    \end{bmatrix}.
\end{equation}
It means that only TE mode propagating to the right will be excited by RCP light. Obviously, the reverse situation will be observed for LCP light, which can excite only left-propagating TE mode.

The same properties can be demonstrated for TM mode.
Although TM mode has non-zero $z$ component of its field we consider its in-plane component since only it interacts with the lattice:

\begin{equation}
    \Tilde{\mathbf{E}}^{\mathrm{TM},\rightleftarrows}_1=
    \begin{bmatrix}
    1\\0
    \end{bmatrix}, \qquad 
    \Tilde{\mathbf{E}}^{\mathrm{TM},\rightleftarrows}_2=
    \begin{bmatrix}
    e^{\pm \mathrm{i} \frac{2\pi}{a_x}\frac{a_x}{4}}\\0
    \end{bmatrix}=
    \begin{bmatrix}
    \pm \mathrm{i}\\0
    \end{bmatrix}.
\end{equation}
These fields result in very similar expression:
\begin{equation}
\begin{bmatrix}
A^{\mathrm{TM},\rightarrow}\\
A^{\mathrm{TM},\leftarrow}
\end{bmatrix}=
    \frac{1}{\omega-\Tilde{\omega}-\alpha/N^{\mathrm{TM}}}\frac{1}{N^{\mathrm{TM}}}
    \begin{bmatrix}1 & -1  &-\mathrm{i} & -\mathrm{i}\\
    1 & -1  &\mathrm{i} & \mathrm{i}
    \end{bmatrix}     \begin{bmatrix}
         \mathbf{E}^0_1\\
         \mathbf{E}^0_2
    \end{bmatrix},
\end{equation}
which corresponds to RCP light exciting left-propagating TM mode and LCP light right-propagating one, which is in accordance with our calculation.


\bibliographystyle{apsrev4-1}

\end{document}